\documentclass[%
 reprint,
nofootinbib,
 amsmath,amssymb,
 aps,
]{revtex4-2}
\usepackage{graphicx}
\usepackage{dcolumn}
\usepackage{bm}
\usepackage{physics}
\usepackage{amsfonts}
\usepackage{footnote}
\usepackage{afterpage}
\usepackage{tikz}
\usepackage{csquotes}
\usepackage[final]{hyperref}
\hypersetup{
	colorlinks=true,       
	linkcolor=blue,        
	citecolor= red,        
	filecolor=magenta,     
	urlcolor=blue         
}
\newcommand{\be}{\begin{equation}}
\newcommand{\ee}{\end{equation}}

\definecolor{purple}{rgb}{1,0,1}
\definecolor{lime}{HTML}{a6CE39} 

\begin{document}
\preprint{APS/123-QED}

\title{Thermal description of braneworld effective theories}
\author{Soham Bhattacharyya\textsuperscript{1}}
\email{soham1316@gmail.com}
\author{Soumitra SenGupta\textsuperscript{1}}
\email{tpssg@iacs.res.in}
\affiliation{$^1$School of Physical Sciences, Indian Association for the Cultivation of Science, Kolkata-700032, India}

\begin{abstract}
Low-energy effective theories provide the natural description of four-dimensional physics in higher-dimensional geometries, where the imprint of the bulk appears as parameters of the lower-dimensional theory. Motivated by the recent progress in the first-order thermodynamic formulation of modified gravity theories, we investigate the thermodynamics of effective theories in braneworld scenarios and thereby the attractor mechanism towards general relativity in such theories. We consider the two-brane Randall-Sundrum model where the low-energy theory on either brane is of scalar-tensor nature with the extra-dimensional radion playing the role of the scalar. We study the thermodynamic implications of a non-vanishing gravitational contribution to the radion potential, and further explore the dynamics in the presence of a bulk stabilizing field.
\end{abstract}

\maketitle


\noindent \textit{Introduction\textemdash} The hypothesis of extra spatial dimensions has become a well-established theoretical idea, wherein our four-dimensional universe is envisioned as a 3-brane embedded in a higher-dimensional spacetime. Such a framework arises naturally in various string-inspired models \cite{Kachru}. Beyond string theory, extra-dimensional scenarios have also been developed as non-supersymmetric alternatives for addressing the well-known fine-tuning or gauge hierarchy problem within the Standard Model of particle physics. Increasingly, it has been recognized that gravity may play a central role in resolving questions that lie beyond the Standard Model. Broadly, extra-dimensional models fall into two main categories: those with large compactification radii \cite{ADD} and those with small compactification radii \cite{RS1}. From a geometric perspective, these dimensions are typically compactified under different topological configurations, with the familiar four-dimensional spacetime emerging in the low-energy limit as an effective theory that retains imprints of the underlying higher-dimensional framework.\\

\noindent  A natural question to ask is- what is the low-energy effective theory on the branes? Since both cosmological and astrophysical solutions occur at energy scales 
much lower than that of the Planck scale, therefore in the effective theory approach
the brane curvature radius $L$ is much large compared to bulk curvature $l$. As a result, perturbation theory can be used with a dimensionless perturbation parameter $\epsilon$ such that $\epsilon = (\frac{l}{L})^2 << 1$.
This method, called the Kanno-Soda gradient approximation scheme \cite{kannosoda}, is a metric-based 
iterative method in which the bulk metric and extrinsic curvature are expanded with increasing orders of $\epsilon$ in perturbation theory. The effective 
Einstein equations on a brane are determined with the solutions of these 
quantities and the junction conditions. When this method is applied to the two-brane Randall-Sundrum (RS) model \cite{RS1}, the RS fine-tuning condition (see Eq.9) is reproduced at the zeroth order when the radion is constant and the two $3$-branes are characterised by opposite 
brane tensions. The effective Einstein equations are then obtained at the first order
incorporating non-zero contributions of the radion field and brane matter and this first order theory has a scalar-tensorial nature - this is also what we obtain by dimensional reduction \cite{chiba}.\\

\noindent In this work, we will confine ourselves mostly to the two-brane Randall-Sundrum model which, along with its variants, have been thoroughly studied over the past two decades \cite{csaki2001,gwvariations,geometry_modulus,SauryaDas_2008,PhysRevD.75.107901,Das_mukherjee,Koley_2009,tanmoy2,banerjeepaul,Banerjeessg, tanmoy,inflation_ssg, Randall_2023, odinstov1, odinstov2, csaki2004tasi,Rubakov_2001,sundrum2005tasi2004lecturesfifth}. The two-brane RS scenario was originally introduced as a geometric resolution of the gauge hierarchy problem, providing an exponential suppression of mass scales on the visible brane without introducing an additional large fundamental hierarchy. This was in stark contrast to large extra-dimensional models, where the hierarchy reappears in the form of a very large compactification radius. Although direct experimental signatures such as Kaluza-Klein (KK) gravitons in collider searches or radion-induced cosmological effects have not yet been detected, it remains plausible that they may just be eluding the present energy scale, since even a $\mathcal{O}(100 TeV)$ KK graviton, which will be beyond the present energy scale, may result from a very little hierarchy introduced between the bulk AdS scale and the radius of the orbifolded compact dimension. From a theoretical standpoint, the RS warped background closely parallels the \enquote{throat} geometries that arise in string theory motivated compactifications. Also as we discussed, RS is one of the fundamental models yielding a scalar-tensor effective theory whose necessity in our further analysis is indispensable.\\

\noindent In its true form, the model has two opposite tension 3-branes embedded in a five-dimensional bulk spacetime. A central challenge in braneworld scenarios is the stabilization of the separation between the two branes - referred to as the modulus or radion. Achieving this requires generating an appropriate potential for the radion field that possesses a stable minimum, consistent with the value prescribed in the RS model, in order to resolve the gauge hierarchy problem. Goldberger and Wise (GW) \cite{GW1} were the first to accomplish this by introducing a bulk scalar field endowed with suitable bulk and brane potentials. They demonstrated that the modulus can be stabilized without resorting to unnatural fine-tuning of parameters. Clearly, upon radion stabilization, the on-brane effective theory which was scalar-tensor (ST) earlier, becomes general relativity (GR) \cite{csakiradion}.\\

\noindent Recently, the first-order thermodynamics of modified gravity theories has gained a lot of interest \cite{giardinothesis,faraoni2,faraoni4,faraoni5,faraoni6,faraoni7,faraoni8,faraoni9,faraoni10,faraoni11,faraoni12,faraoni13,faraoni14,faraoni15,faraoni16,faraoni18,horndeskifaraoni1,horndeskifaraoni2,faraoni17horndeski}. The primary idea is- if the effective fluid which appears in the modified Einstein equation of the corresponding gravity theory satisfies the Eckart-Fourier constitutive equation for dissipative fluids \cite{eckart}, then a \enquote{temperature} of the gravity theory can be defined consistently. Miraculously, one of such theories is scalar-tensor gravity, which has consequently led to a lot of exploration \cite{giardinothesis,faraoni2,faraoni4,faraoni5,faraoni6,faraoni7,faraoni8,faraoni9,faraoni10,faraoni11,faraoni12,faraoni13,faraoni14,faraoni15,faraoni16,faraoni18}. More generally, Horndeski theories pass the same test \cite{horndeskifaraoni1,horndeskifaraoni2,faraoni17horndeski}. Much recently, Ref.\cite{faraoni1} uses the first-order thermodynamic interpretation to formalize the attractor-to-GR mechanism in scalar-tensor theories. This was first proposed by Damour and Nordverdt \cite{damour1,damour2} and was subsequently plagued by a repellor mechanism \cite{repellor} which only led to increasing complications in understanding the mechanism. Ref. \cite{faraoni1} however provided a fresh perspective on the issue - it linked the relaxation of the effective fluid to its \enquote{zero temperature} equilibrium state to the attractor mechanism kicking in and taking scalar-tensor gravity towards GR, and vice versa. This formalism thus introduces the concept of a “gravitational temperature” along with an explicit equation governing the system’s evolution toward, or away from, GR equilibrium, making it particularly well-suited for studying the attractor-to-GR mechanism in cosmology.\\

\noindent Given the scalar-tensorial nature of effective field theories in higher dimensional warped spacetimes, we believe that it should be worthwhile to explore these braneworld effective theories in the new light of first-order scalar-tensor thermodynamics, with an emphasis on the attractor-to-GR mechanism. We start off with a quick review of the first-order thermodynamics of ST gravity.\\


\noindent \textit{First-order ST thermodynamics and the attractor-to-GR mechanism\textemdash} We take first-generation ST gravity of the form \cite{faraoni1,faraoni9,faraoni10,faraoni11},
\be 
S = \int \frac{d^4 x}{16\pi} \sqrt{-g} \left[ \phi R 
-\frac{\omega(\phi)}{\phi} \, \nabla^a \phi \nabla_a \phi -V \right] + 
S_m  
\ee 
with $g_{ab}$ being the metric and $R$ the Ricci scalar. The Brans-Dicke (BD) function $\omega>-3/2$ to avoid $\phi$ being  a 
phantom field, and $S_m$ is the matter action.\\ 

\noindent With a timelike and future-oriented gradient $\nabla^a\phi$,  the effective energy-momentum tensor $T_{ab}^{(\phi)}$ of $\phi$ is equivalent to that of an effective dissipative fluid with 4-velocity 
$u^a := \nabla^a \phi /\sqrt{-\nabla^e \phi \nabla_e\phi}$. The stress tensor of a dissipative imperfect fluid of 4-velocity $u^a$, in its full generality, can be written as
\be
T_{ab} =\rho u_a u_b 
+P h_{ab} +\pi_{ab} + q_a u_b + q_b u_a
\label{eq:imperfect}
\ee
\\
where $\rho$, $P$, $\pi_{ab}$ and $q_a$ represent an effective energy density, effective isotropic pressure, effective anisotropic stress tensor and effective heat flux density respectively. $h_{ab} = g_{ab}+u_au_b$ (describes hypersurface perpendicular to $u_a$). The fluid quantities associated with $T^{(\phi)}_{ab}$ are denoted with a $(\phi)$ superscript. The observation that truly led to this alternate description of ST gravity is that this effective fluid obeys the Eckart-Fourier constitutive law for dissipative fluids \cite{eckart} 
$ 
q_a = -{\cal K} \left( \nabla_a {\cal T} + {\cal T} \dot{u}_a \right)$ where ${\cal K}$ is the thermal conductivity, ${\cal 
T}$ is the temperature, and $\dot{u}^a = u^c \nabla_c u^a$ is the fluid acceleration. A pedagogical calculation infact yields $q_a^{(\phi)} =-{\cal KT} \dot{u}_a$ where  
\cite{faraoni1,faraoni9,faraoni10,faraoni11} 
\be  
{\cal KT} = \frac{ 
\sqrt{-\nabla^c\phi \nabla_c\phi} }{ 8\pi \phi} \, \label{KTdefinition} 
\ee 
As we observe, ST gravity reduces to GR in the limit $\phi \to$~constant i.e. ${\cal KT} \to 0$. 
The convergence (divergence) of ST gravity to (from) GR is described by \cite{faraoni1,faraoni9,faraoni10,faraoni11} 
\be 
\frac{d\left( {\cal KT}\right)}{d\tau} = 8\pi \left( {\cal KT}\right)^2 
-\Theta {\cal KT} +\frac{ \Box\phi}{8\pi \phi} \,  \label{evolution_general} 
\ee 
where 
$\tau$ is the proper time along the $\phi$ fluid flow and 
$\Theta := \nabla_c u^c$ is its expansion scalar. Incorporating the scalar equation of motion to substitute for $\square\phi$, we get
\be
\begin{split}
\frac{d \left( {\cal KT}\right)}{d\tau} = 8\pi \left( {\cal KT}\right)^2 
-\Theta {\cal KT} + \frac{ T^\mathrm{(m)} }{\left(  2\omega + 3 \right) \phi}\\ +\frac{1}{8\pi 
\left( 2\omega + 3 \right)} \left(  V' -\frac{2V}{\phi} -\frac{\omega'}{\phi} \, 
\nabla^c\phi \nabla_c\phi \right) \,
\end{split} 
\ee  
Using the definition of $\cal KT$ from Eq.(3) , the above equation simplifies as \cite{faraoni1},
\be
\begin{split}
\frac{d \left( {\cal KT}\right)}{d\tau} &= 8\pi \left[ 1+  \frac{\phi}{2} \, \frac{d\ln(2\omega+3)}{d\phi} \right] \left( {\cal KT}\right)^2 -\Theta \, {\cal KT}\\ &\quad +\frac{ T^{(m)} }{\left( 2\omega + 3 \right) \phi} +\frac{V'-2V/\phi}{8\pi \left( 2\omega + 3 \right)} \,
\label{BOH}
\end{split}
\ee
Gravity  is ``heated'' ($d\left( {\cal KT}\right)/d\tau>0$) by positive terms in the  right-hand side and is ``cooled'' ($d\left({\cal  KT}  \right)/d\tau<0$) by negative ones. The positive (negative) contributions therefore make it easier for gravity to diverge away (converge to) GR.\\

\noindent \textit{Braneworld EFTs\textemdash} We briefly review the background metric found by Randall and Sundrum 
\cite{RS1} to introduce our notations. The set up is five-dimensional Einstein gravity
with a cosmological constant with two 3-branes located in the orbifold
$S^1/{\bf Z}_2$ at the fifth dimensional coordinate $z=0$ (Planck brane) and $z=r_c$ (TeV brane). The action is
\be
\begin{split}
S=2\int d^4x\int_{0}^{r_c} dz\sqrt{-g_5}\left[
{2M^3}R_5-\Lambda_b\right]\\
-T_{hid}\int d^4x\sqrt{-g_{(+)}}
-T_{vis}\int d^4x\sqrt{-g_{(-)}},
\end{split}
\ee
where $M$ is the five-dimensional Planck mass defined in term of the 
five-dimensional gravitational constant $G_5$ as $M^{-3}=8\pi G_5$. 
$T_{(hid,vis)}$ and $g_{(\pm)}$ are the brane tension and the
induced metric on the brane, respectively. 
Randall and Sundrum showed that there exists a solution that
respects four-dimensional Poincar\'e invariance \cite{RS1}:
\be
ds^2=e^{-2k |z|}\eta_{\mu\nu}dx^{\mu}dx^{\nu}+dz^2,
\label{back}
\ee
only if $\Lambda_b, T_{hid}$ and  $T_{vis}$ are related as
\be
\Lambda_b=-24M^3k^2, T_{hid}=-T_{vis}= -\Lambda_b/k.
\label{relation}
\ee
\noindent 
To include massless
gravitational degrees of freedom (the zero modes about the background
spacetime Eq.(\ref{back})), we replace the Minkowski metric
$\eta_{\mu\nu}$ with a general metric $g_{\mu\nu}(x)$ and 
$z$-direction length $r_c$ with a modulus field $T(x)$ \cite{RS1,GW2}: 
\be
ds^2=e^{-2kT(x)|z|}g_{\mu\nu}(x)dx^{\mu}dx^{\nu}+T(x)^2dz^2
\label{metricnaive}
\ee
The induced metric on the positive (or negative) tension brane is thus  
$g_{(+)\mu\nu}=g_{\mu\nu}$ (or $ g_{(-)\mu\nu}= e^{-2
kr_cT}g_{\mu\nu}$), respectively. 
Since we impose ${\bf Z}_2$ symmetry ($z\leftrightarrow -z$), 
massless vector fluctuations associated with the off-diagonal part of 
the metric are absent \cite{RS1,GW2}. The low-energy effective theory can now be found by simple dimensional reduction (or by gradient expansion method as in Ref.\cite{kannosoda}).\\

\noindent What we get is that the effective 4-D theory on either brane is of the Brans-Dicke form with the following action
\be
S=\int d^4x\sqrt{-g}{1\over
16\pi}\left(\Phi_{\pm}R-{\omega(\Phi_{\pm})\over \Phi_{\pm}}(\nabla
\Phi_{\pm})^2\right)
\ee
The Brans-Dicke scalar field (or the inverse of the effective gravitational constant) is 
\be
\begin{split}
{1\over G_{\rm eff}^{\pm}}=\Phi_{\pm}=
{64\pi M^3\over k}e^{\mp kr_cT}\sinh(
kr_cT)
\end{split}
\ee
while the corresponding Brans-Dicke function is
\be
\omega_{\pm}(T)=\pm 3e^{\pm kr_cT}\sinh(kr_cT)
\ee
with $0< \omega_+ < \infty$ for the Planck brane and
$-3/2< \omega_- <0$ for the TeV brane.\\

\noindent The Planck (TeV) brane is the positive (negative) tension brane. We denote the Brans-Dicke scalar on the brane by $\Phi_+$ ($\Phi_-$) and in terms of the scalar, the Brans-Dicke function can be written as
\be
\omega_+ = \frac{3}{2} \frac{\Phi_+}{\alpha - \Phi_+},\; \omega_- = -\frac{3}{2} \frac{\Phi_-}{\alpha + \Phi_-}
\ee
where $\alpha = 32\pi M^3/k$. Consequently, its derivative would be
\be
\omega_+' = \frac{3}{2}\frac{\alpha}{(\alpha - \Phi_+)^2},\; \omega_-' = -\frac{3}{2}\frac{\alpha}{(\alpha + \Phi_-)^2}
\ee
Thus, we already naively see that the last term in Eq.(5) whose sign solely depends on $\omega_\pm'$ is (a) [on the Planck brane] strictly positive and hence heats up gravity whereas (b) [on the TeV brane] is strictly negative and hence cools down gravity.\\

\noindent If we were to write Eq.(6) in the present context, it would take the following form upon substituting the Brans-Dicke function,
\be
\frac{d({\cal KT})}{d\tau} = 8\pi (1+\frac{\omega_\pm}{3}) ({\cal KT})^2  - \Theta {\cal KT}
\ee
with
\be
(1+\frac{\omega_\pm}{3}) = \frac{1+ e^{\pm 2kr_cT}}{2}
\ee
\noindent As we can see from Eq.(16), there are only two solutions with ${\cal KT}$ = constant ($\frac{d({\cal KT})}{d\tau}=0$). These are ${\cal KT} = 0$ and ${\cal KT} = \Theta/8\pi (1+\frac{\omega_{\pm}}{3})$. Following the analysis of Ref.\cite{faraoni1}, we demonstrate this scenario on the $(\Theta,{\cal KT})$ plane (see Fig.1), albeit only the first quadrant. The quadrants where ${\cal KT} < 0$ are discarded by definition. We also need not worry about the ($\Theta<0,{\cal KT}>0$) quadrant as Eq.(16) becomes: $\frac{d({\cal KT})}{d\tau} = 8\pi (1+\frac{\omega_\pm}{3}) ({\cal KT})^2  + |\Theta| {\cal KT}$. $(1+\frac{\omega_\pm}{3})$ being positive, this quadrant naturally takes away all solutions away from GR. The ${\cal KT} = \Theta/8\pi (1+\frac{\omega_{\pm}}{3})$ solution is identified as the critical curve - solutions lying below (above) it converge to (diverge from) GR. As the Brans-Dicke functions for both branes depend on the radion field, the critical curves are not straight lines. The comparative positions of critical curves on either brane can easily be seen from Eq.(17) - since $\frac{8\pi {\cal KT}}{\Theta} = \frac{1}{1+\frac{\omega_+}{3}}$, the Planck brane critical curve is below $8\pi {\cal KT} = \Theta$ line (this is representative of electrovacuum Brans-Dicke \cite{faraoni1}) as $(1+\frac{\omega_+}{3}) > 1$ whereas the TeV brane critical curve is above the same as $\frac{1}{2}<(1+\frac{\omega_-}{3})< 1$. Owing to the relative positions of the critical curves (see Fig.1), the TeV brane allows a larger class of solutions to flow towards GR than the Planck brane - in this sense, the TeV brane acts as a better \enquote{cooler} for ST theories than the Planck brane. We should remember that owing to the dependence of both the BD scalars on the radion field (see Eq.12), the BD scalar on one brane going to a constant value automatically implies the same for the other brane's BD scalar - hence, the signature of $\frac{d(\cal KT)}{d\tau}$ is identical on either brane. Thus, the apparent better cooling nature of the TeV brane stems from the parametric dependence of the BD scalars on the radion field as
\begin{equation}
    e^{2k r_c T} = \frac{\alpha}{\alpha - \Phi_+} = 1+\frac{\Phi_-}{\alpha}
\end{equation}
As one can observe, $\Phi_-$ is unbounded above whereas $\Phi_+$ can only go up to $\alpha$, thereby explaining the disparity.\\

\noindent An immediate question that arises is whether the critical curve on one brane can have the exact same profile as that on the other brane when plotted on the $(\Theta,{\cal KT})$ plane. A quick calculation shows that $\omega_+ = \omega_-$ only when $T=0$ i.e. the two branes overlap effectively reproducing the single brane Randall-Sundrum model \cite{RS2}.\\

\noindent In case of an FLRW geometry on the branes with $a(t)$ being the scale factor, $\Theta = 3H$ where $H=\frac{\dot{a}}{a}$. For the attractor mechanism to kick in, we need to have
$8\pi {\cal KT} < \frac{\Theta}{1+\frac{\omega_\pm}{3}}$.
Using the definition of ${\cal KT}$ from Eq.(3), we get $\frac{|\dot{\Phi}|}{\Phi} < \frac{3H}{1+\frac{\omega_\pm}{3}}$.
In a contracting universe ($H<0$), for the attractor mechanism to work, $(1+\frac{\omega_\pm}{3})$ has to be negative which is forbidden - hence, no attractor mechanism in such a universe. On the contrary, in an expanding universe ($H>0$), the attractor mechanism can indeed take place on both the branes provided that we satisfy

\be
\tau_{\Phi} > (1+ \frac{\omega_\pm}{3}) \frac{\tau_H}{3}
\ee

\noindent where the scale of variation of the field $\tau_{\Phi} = \frac{|\dot{\Phi}|}{\Phi}$ and $\tau_H = \frac{1}{H}$. As Ref.\cite{faraoni1} highlights, this is essentially a competition between the scalar mode and the two massless spin-2 GR modes. As $\frac{1}{2}<(1+\frac{\omega_-}{3})< 1$, it diminishes the $\frac{\tau_H}{3}$ contribution, helping the attractor mechanism on the TeV brane whereas the Planck brane witnesses the exact opposite situation as $(1+\frac{\omega_+}{3})>1$ increases the right hand contribution of Eq.(19) thereby making the attractor mechanism less likely.\\

\noindent Braneworld models are commonly associated with ekpyrotic and cyclic cosmology \cite{LEHNERS2008223}. The model postulates an attractive force between the two branes that causes them to slowly approach one another, leading to an ekpyrotic phase in which the universe undergoes a slow contracting stage that flattens the branes to a high degree aided by the on-brane ST theory. Eventually, the branes collide, and from the perspective of observers on a brane this collision appears as the Big Bang. After the collision, the brane separation nearly stabilizes, but a small residual attraction remains, acting like quintessence and providing the observed dark energy, and over very long timescales the attraction strengthens again, leading to another ekpyrotic phase, another collision, and thus a cyclic cosmology with repeated periods of contraction, collision, and expansion. In such cosmologies, the thermal description for an FLRW geometry tells us that for the contracting phase ($H<0$), no attractor-to-GR mechanism is possible but it may again get activated in the subsequent expanding phase ($H>0$). Similar analysis can be carried out for more general geometries.\\

\begin{figure*}
    \centering
    \includegraphics[width=0.9\linewidth]{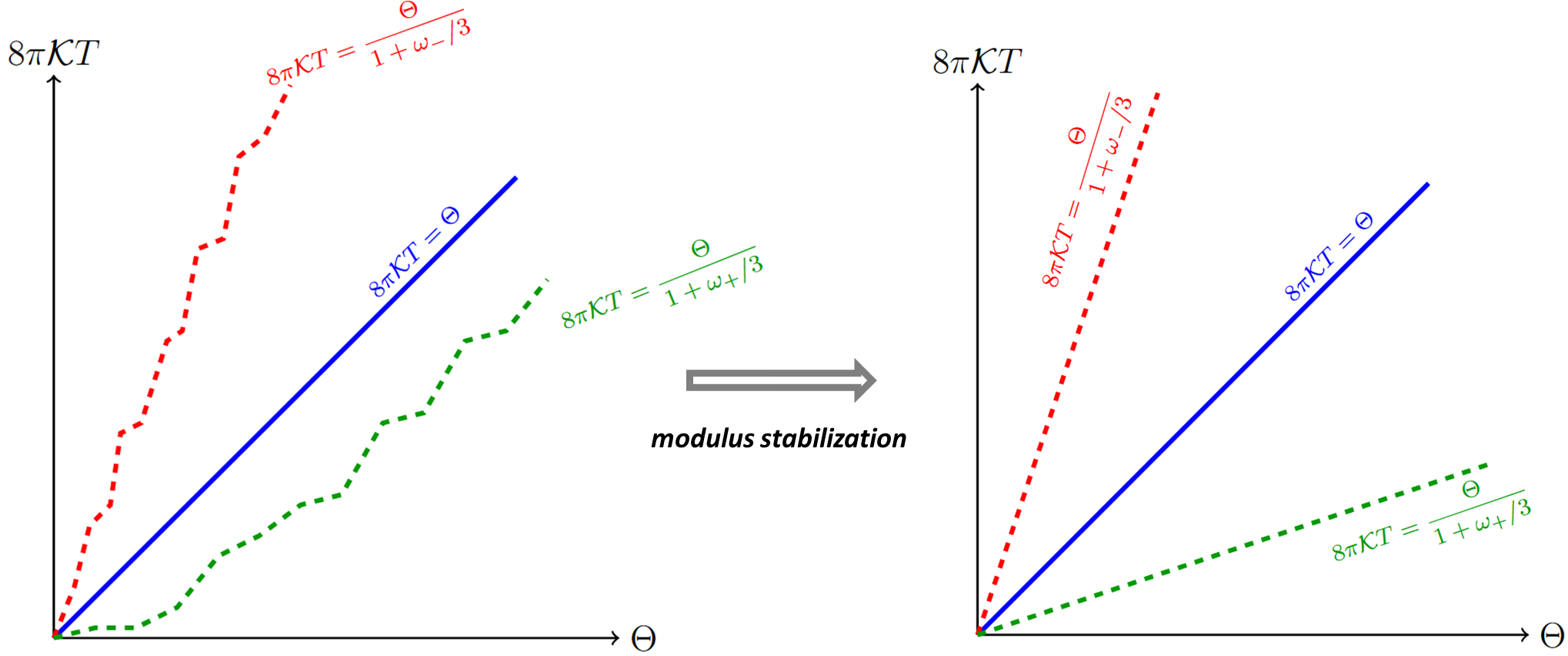}
    \caption{A schematic representation of the critical curves in three specific theories - electrovacuum Brans-Dicke (blue) [see Ref.\cite{faraoni1}], TeV brane Brans-Dicke (red) and Planck brane Brans-Dicke (green). For each of the theories, solutions starting above the critical curve
deviate forever from GR, those starting below converge to
GR. The critical curves cannot be crossed. \textit{On left} - without a stabilizing field. \textit{On right} - with a stabilizing field - dynamics about the stabilized radion.}
    \label{fig:placeholder}
\end{figure*}

\noindent \textit{Effect of the gravitational contribution to radion potential\textemdash} In getting to Eq. (11) as the effective four-dimensional theory, we assumed that the original RS fine-tunings would still hold (Eq.9). This made the gravitational contribution to radion potential vanish. However, we could take the brane tensions to be detuned \cite{GW2,Chacko1,sbssg1}, and then the non-vanishing radion potential would alter the thermodynamics of the on-brane scalar-tensor theory. Defining $\varphi\equiv Ae^{-kr_c T(x)}$, where $A=\sqrt{24M^3/k}$ to be the canonical radion field, one obtains the following form for the gravitational contribution to radion potential
\begin{equation}\label{9}
\begin{split}
    V_{GR}(\varphi)= \frac{k^4}{A^4}\left(\tau\:\varphi^4+\Lambda_{4D}\right)
\end{split}
\end{equation}
where $\Lambda_{4D}=\Big(T_{hid}+\frac{\Lambda_b}{k}\Big)/k^4$ is the $4D$ cosmological constant on the Planck brane, and $\tau=\Big(T_{vis}-\frac{\Lambda_b}{k}\Big)/k^4$.\\

\noindent \textit{On Planck brane} : If we were to recast the potential in Eq.(20) in terms of the BD scalar, it would turn out to be
\be
V_{GR}(\Phi_+) = \frac{k^4}{A^4}[(1-\frac{\Phi_+}{\alpha})^2 A^4 + \Lambda_{4D}]
\ee
Calculating $V'-\frac{2V}{\Phi_+}$ where the prime represents derivative with respect to $\Phi_+$ , we get
\be
V'-\frac{2V}{\Phi_+} = \frac{2k^4}{A^4}[\tau A^4 (\frac{1}{\alpha}- \frac{1}{\Phi_+}) - \frac{\Lambda_{4D}}{\Phi_+}]
\ee
Let us break it down to two regimes: (a) $\Lambda_{4D} \rightarrow 0$: the 4D cosmological constant tuned to zero implies an almost flat Planck brane. This reduces the above combination to just the first term. $\Phi_+ < \alpha$ implies $(\frac{1}{\alpha}- \frac{1}{\Phi_+})<0$. Thus, the contribution now solely depends on the detuning of the visible brane tension. If $T_{vis}>\Lambda_b/k$, then the contribution is negative overall and it cools down gravity and vice versa. (b) $\tau \rightarrow 0$: This gets rid of the first term and the signature of the contribution depends only on the 4D cosmological constant i.e. detuning of the hidden brane tension. If $T_{hid}>-\Lambda_b/k$, then the contribution is negative implying a cooling down of gravity - and vice versa. The only thing we could decipher without pertaining to any regime is the following: if $T_{vis}>\Lambda_b/k$ and $T_{hid}>-\Lambda_b/k$ can be ensured simultaneously (which is certainly greater detuning), then the net contribution is negative - hence, making it easier for the theory to tend towards GR. But this in turn means that $T_{hid}+T_{vis}>0$. Recalling the Gibbons-Kallosh-Linde geometric consistency condition \cite{branesum1,branesum2}, 
\begin{equation}
     T_{hid} + T_{vis}
= - M^3 R_p\oint e^{-2 kr_cT(x)}
\end{equation}
where \enquote{$\oint$} is the integration over the internal compact space and $R_p$ is the Ricci scalar for the Planck brane metric. Thus, we can have the $T_{hid}+T_{vis}>0$ only if the Planck brane geometry is of the anti-de Sitter type ($R_p<0$). Similarly, heating the theory would require a de-Sitter Planck brane geometry ($R_p>0$).\\

\noindent \textit{On TeV brane} : The TeV brane action is obtained from the Planck brane action by the conformal transformation $ g_{(-)\mu\nu}= e^{-2
kr_cT}g_{(+)\mu\nu}$. Hence, the TeV brane potential is obtained by multiplying Eq.(20) with $e^{4kr_cT}$. In terms of the BD scalar, it would then be
\begin{equation*}
V_{GR}(\Phi_-) = \frac{k^4}{A^4} (\frac{\Phi_-+\alpha}{\alpha})^2[(1-\frac{\Phi_-}{\alpha})^2 A^4 + \Lambda_{4D}]
\end{equation*}
\begin{equation}
\implies V_{GR}(\Phi_-) = \frac{k^4}{A^4}[\Lambda'_{4D}+\tau'(1+\frac{\Phi_-}{\alpha})^2]
\end{equation}
where $\Lambda'_{4D} = A^4 \tau$ is the 4D cosmological constant for the TeV brane and $\tau' = \Lambda_{4D}$.
Calculating $V'-\frac{2V}{\Phi_-}$ where the prime denotes derivative with respect to $\Phi_-$, we get
\be
\begin{split}
V'-\frac{2V}{\Phi_-} = -\frac{2k^4}{A^4}[\tau'(\frac{1}{\alpha}+\frac{1}{\Phi_-})+\frac{\Lambda'_{4D}}{\Phi_-}]
\end{split}
\ee
One can observe Eq.(25) to reach the same conclusions as in the Planck brane case i.e. in the different regimes, the same detunings yield the same signatures for the contribution \footnote{The physical meaning of $\tau \rightarrow 0$ is $\Lambda'_{4D} \rightarrow 0$ i.e. an almost flat TeV brane.}. This is in fact an outcome of the stronger result we obtain upon conformal tansformation,
\begin{equation}
\frac{dV}{d\Phi_+} - \frac{2V}{\Phi_+} = \frac{dV}{d\Phi_-} - \frac{2V}{\Phi_-}
\end{equation}


\noindent A crucial observation is that in both the regimes (\enquote{almost flat Planck brane} and \enquote{almost flat TeV brane}), non-dynamical detunings of the brane tensions entirely determine the signature for the radion potential contribution to the ${\cal KT}$ dynamics of the gravity theory (whether it \enquote{cools} or \enquote{heats} gravity) on either brane. For the $\Lambda_{4D} \rightarrow 0$ regime, the signature of $(T_{vis}-\Lambda_b/k)$ does the job; and for the $\Lambda'_{4D} \rightarrow 0$ regime, the signature of $(T_{hid}+\Lambda_b/k)$ plays the equivalent role.\\

\noindent \textit{Including a bulk stabilizing field\textemdash} Originally, Ref.\cite{RS1} set the potential in Eq.(20) to zero through fine tunings of the brane tensions i.e. $T_{vis} = -T_{hid} = \Lambda_b/k$. However, then the radion field could take up any possible value. To fix this, one needed a stabilization mechanism that fixes the interbrane separation and makes the radion acquire a mass. Such a mechanism was proposed by 
Goldberger and Wise \cite{GW1}. In this construction, a 
massive $5D$ field $\Phi$ is sourced at the boundaries and acquires a 
VEV whose value depends on the location in the extra dimension. After 
integrating over the extra dimension, this generates a radion potential in the low energy effective theory. In the original GW 
construction, the quartic potential for the radion from the gravity sector was still kept tuned to zero, and only the dynamics of the scalar field $\Phi$ 
contributed to the radion potential.\\

\noindent One should note that in the absence of a stabilization mechanism, the radion remains massless but not tachyonic, reflecting a flat direction rather than an instability of the background \cite{RS1,chiba,GW1,GW2}. The transverse traceless graviton perturbations reduce to a Schrodinger type eigenvalue problem with the usual \enquote{volcano} potential, yielding a normalizable zero-mode reproducing four-dimensional gravity together with a KK tower satisfying $m^2 > 0$ \cite{RS1,RS2,csaki2004tasi,Rubakov_2001}. Thus, the absence of a stabilizing field does not by itself generate geometric pathologies or tachyonic KK instabilities - instead, the inter-brane separation remains dynamically undetermined.\\

\begin{figure*}
    \centering
    \includegraphics[width=0.9\linewidth]{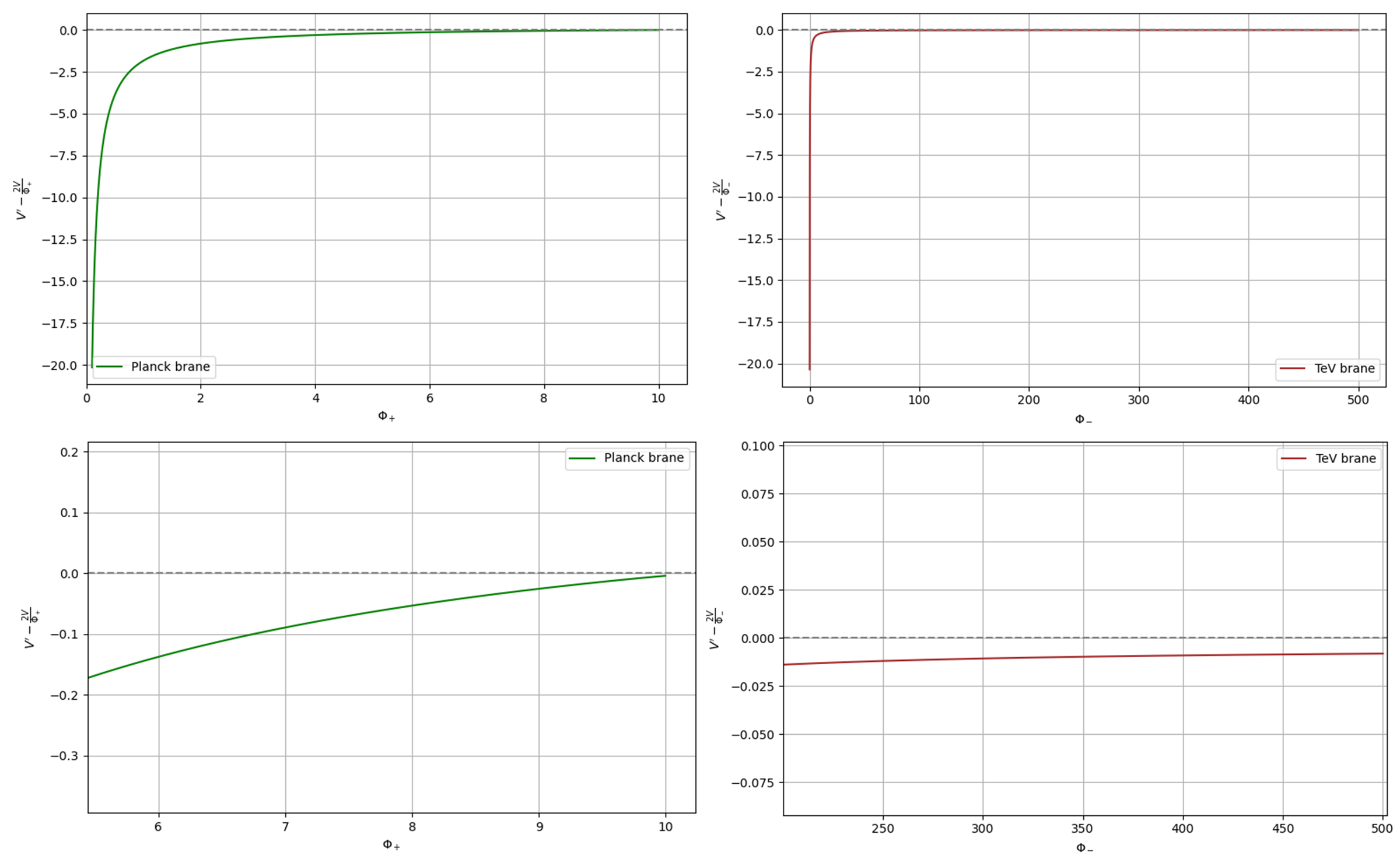}
    \caption{$(V'-\frac{2V}{\Phi_\pm})$ vs $\Phi_\pm$ on Planck brane (in green) and on TeV brane (in brown) with parameter values $v_h/v_v=1.5$, $\alpha=10$ and $\epsilon = m^2/4k^2 = 0.01$ for a positive mass squared bulk scalar field. Bottom figures zoom in at large $\Phi_\pm$ values.}
\end{figure*}

\noindent We add to the original RS action a scalar field $\Phi$ with the following bulk action
\begin{equation}\label{10}
S_b=\int d^4 x\int_{0}^{r_c} dz \sqrt{G} \left(G^{AB}\partial_A \Phi \partial_B \Phi - m^2 \Phi^2\right)
\end{equation}
where $G_{AB}$ with $A,B=\mu, z$ is given by Eq.(10).  We also include interaction terms on the hidden and visible branes (at $z=0$ and $z=r_c$ respectively) given by
\begin{equation}\label{11}
S_h = -\int d^4 x \sqrt{-g_h}\lambda_h \left(\Phi^2 - v_h^2\right)^2
\end{equation}
and
\begin{equation}\label{12}
S_v = -\int d^4 x \sqrt{-g_v}\lambda_v \left(\Phi^2 - v_v^2\right)^2
\end{equation}
We get a $z$-dependent vacuum expectation value $\Phi(z)$ which is determined classically. Plugging this solution back into the scalar field action and integrating over $z$ yields an effective four-dimensional potential $V_\Phi$ for $T(x)$ (in the large coupling limit),

\begin{equation}\label{17}
\begin{split}
V_{GW}(T)= k\epsilon v_h^2 + 4ke^{-4kr_cT(x)}(v_v - v_h e^{-\epsilon kr_cT(x)})^2\left(1+\frac{\epsilon}{4}\right)\\ - k\epsilon v_h e^{-(4+\epsilon)kr_cT(x)}(2 v_v - v_h e^{-\epsilon kr_cT(x)})
\end{split}
\end{equation}
with $\epsilon = m^2/4k^2$ - terms of order $\epsilon^2$ are neglected. 
Ignoring terms proportional to $\epsilon$, this potential has a minimum at 
\begin{equation}\label{18}
kr_cT =  \frac{4k^2}{m^2} \ln\left[\frac{v_h}{v_v}\right]
\end{equation}
Using $v_h/v_v=1.5$ and $m/k = 0.2$ in Eq.(\ref{18}) yields $kr_cT\approx 36$ and as one can see, no unnatural fine-tuning of parameters is required to solve the hierarchy problem \cite{GW1}. Since the modulus gets stabilized to a fixed value, we can clearly see that the Brans-Dicke fields on both the branes which depend on the modulus field (Eq.12) become constants. As a result, the theory on either brane eventually turns out to be GR (${\cal KT} = 0$). The question is if we can witness the same from the thermal viewpoint i.e. the dynamical equation for ${\cal KT}$ without adhering to any particular geometry (like FLRW) on the brane. Accordingly, we plotted $V'-\frac{2V}{\Phi_\pm}$ using Eqs. (12) and (30). The numerics were performed for the choice of parameters below Eq.(31) which do yield a value of the stabilized modulus deemed suitable for addressing the hierarchy problem. The result we get is the following: on both the branes, the signature of the $V'-\frac{2V}{\Phi_\pm}$ contribution is negative till large values of $\Phi_\pm$ (on the Plack brane, the $\Phi_+$ axis is complete since $T\rightarrow \infty$ implies $\Phi_+ \rightarrow \alpha$), hence cooling down the gravity theory on either brane (see Fig.2) - which is precisely what we expect to occur in the presence of a stabilizing field.\\

\noindent It is well known that a stabilizing field with negative mass term fails to stabilize the radion \cite{sbssg1,GW1} despite satisfying the Breitenlohner-Freedman bound. \footnote{In AdS space which is the nature of the bulk in Randall-Sundrum, the stability of the stabilizing field only requires satisfying the Breitenlohner-Freedman bound \cite{1982}: $m^2/k^2+4>0$ where $m^2$ is the mass squared of the GW field and $k$ is the AdS curvature scale.} As the modulus never settles to a fixed value, we do not expect the gravity on either brane to flow towards GR. The same numerics were performed and the results show that the signature of $V'-\frac{2V}{\Phi_\pm}$ on either brane eventually becomes positive for large enough values of $\Phi_\pm$. As the bulk field fails to stabilize the radion, the radion blows up (as the radion potential is ever decreasing) which indicates $\Phi_+$ reaching $\alpha$ and $\Phi_- \rightarrow \infty$. Hence, the presence of such a bulk field drives the increment of the radion field, which upon crossing a threshold renders the signature of $V'-\frac{2V}{\Phi_\pm}$ positive (see Fig.3) - this drives the gravity theory on either brane away from GR \footnote{One may worry that the TeV brane scalar is unbounded - hence commenting on the contribution's signature throughout is unjustified. However, this is not the case as Eq.(26) tells us that the contribution is identical on both branes and the entire range of the Planck brane BD scalar has been explored, providing a full knowledge of the TeV brane contribution.}\\
\begin{figure*}
    \centering
    \includegraphics[width=0.9\linewidth]{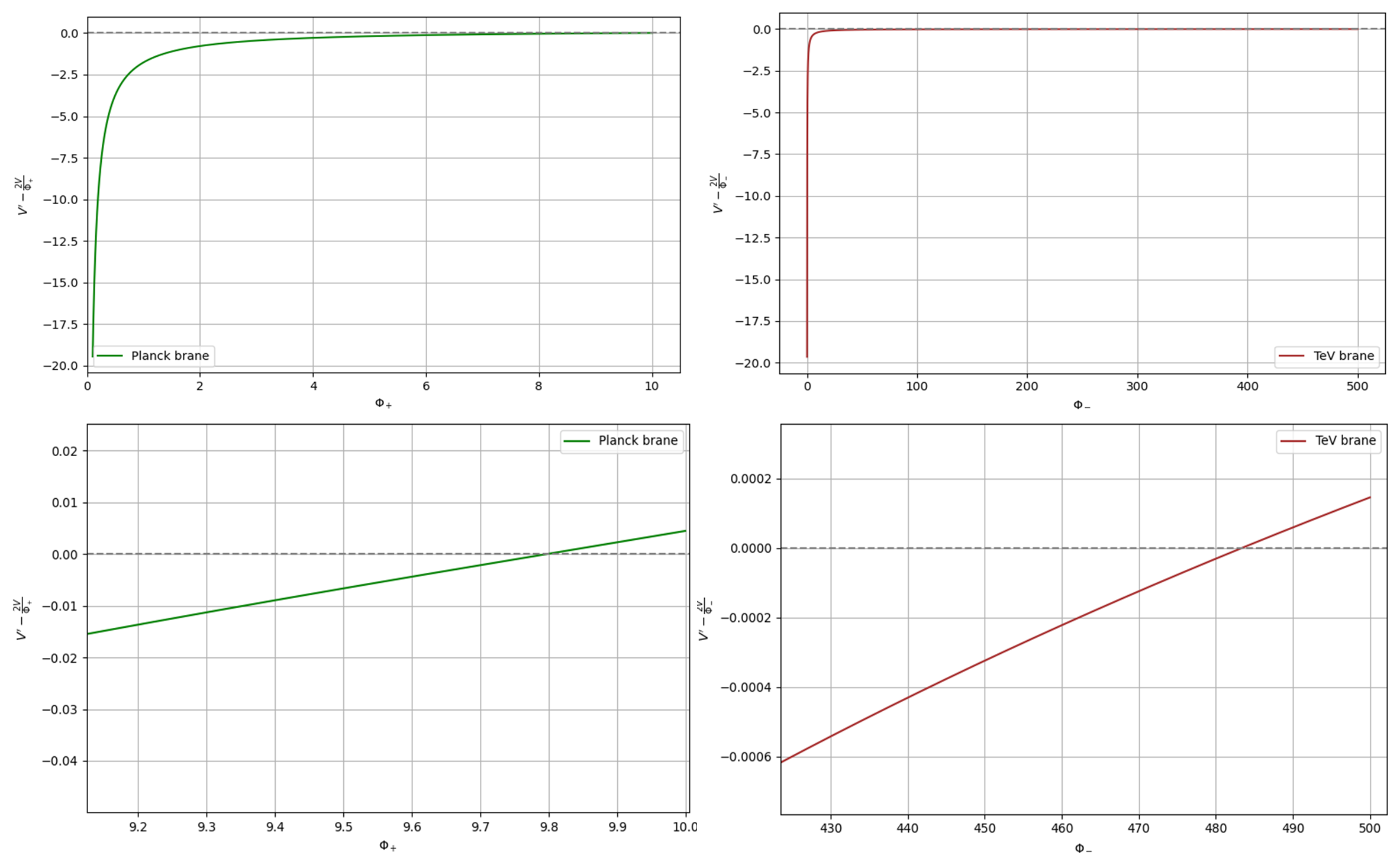}
    \caption{$(V'-\frac{2V}{\Phi_\pm})$ vs $\Phi_\pm$ on Planck brane (in green) and on TeV brane (in brown) with parameter values $v_h/v_v=1.5$, $\alpha=10$ and $\epsilon = m^2/4k^2 = -0.01$ for a negative mass squared bulk scalar field. Bottom figures zoom in at large $\Phi_\pm$ values.}
\end{figure*}

\noindent What exactly does radion stabilization do to the critical curves in the $(\Theta,{\cal KT})$ plot? We discussed that radion settles to a stable minimum of the radion potential with any residue dynamics being fluctuations about this stable minimum. As previously mentioned, the on-brane theory would eventually converge to GR. Thus, modulus stabilization effectively does two jobs - first, it straightens the critical curves - as the slopes depend on Brans-Dicke functions which in turn only depend on the radion field; they become almost constant - and second, it takes all possible cosmolgical solutions to the lower half of the corresponding critical line, ensuring the convergence of the theory to GR (see Fig.1). One may wonder that upon including a stabilizing field, we would get the radion potential contribution on the right side of Eq.(16) which would hinder the critical curves from being straight lines. But since the dynamics is now restricted to around the stable minimum, the radion potential is effectively $V(\Phi_\pm) \propto \Phi_\pm^2$. This makes the contribution $(V'-\frac{2V}{\Phi_\pm})$ vanish and hence, we still retain the form of Eq.(16).\\

\noindent We have two different ST theories on two branes, but what makes things interesting is that the metrics on these branes are conformally related and the BD scalars on the branes are connected by the same conformal factor i.e. $g_{(-)\mu\nu} = e^{-2kr_cT} g_{(+)\mu\nu}$ and $\Phi_- = e^{2kr_cT}\Phi_+$. On modulus stabilization, we develop a radion potential with a minimum and the modulus value hovers about this VEV. Hence, the conformal factors are treated as constants. Then, by the definitions of fluid velocity and fluid temperature (Eq.5), we get to the following result:

\be
\frac{d}{d\tau_-}({\cal K_-T_-}) = e^{2kr_cT}\frac{d}{d\tau_+}({\cal K_+T_+})
\ee

\noindent This essentially tells us that the ${\cal KT}$ dynamics on either brane is equivalent upto a conformal factor - if we know the dynamics on one brane, we know the dynamics on the other. At a glance, this may seem like a non-local effect, but it is truly an artefact of the facet that the Brans-Dicke field on either brane depends on the radion field, which governs the inter-brane separation. In the general scenario, one could do the conformal transformation explicitly to get the dynamics on TeV brane but then the relation would not be this direct.\\

\noindent \textit{On other braneworld theories\textemdash} There are several braneworld theories whose low-energy effective theories are of the scalar-tensor nature: RS with non-flat branes \cite{SauryaDas_2008,nonflateffective}, f(R) braneworlds \cite{f(R)braneworld}, etc. - but these ST theories have non-canonical couplings for the kinetic terms and are often extremely complicated. A particular simplification happens when the kinetic coupling of such a theory vanishes as it can then be rewritten as an $f(R)$ theory. In the context of first-order thermodynamics of ST gravity, $f(R)$ theories are inherently important because of their correspondence to ST theories. In the Jordan frame $f(R)$, $\Psi = f'(R)$ and $V(\Psi) = Rf'(R) - f(R)$. where $\Psi$ is the Jordan frame scalar. In this context, Eq. (6) takes the form,
\be
\begin{split}
\frac{d \left( {\cal KT}\right)}{d\tau} = 8\pi \left( {\cal KT}\right)^2 -\Theta \, {\cal KT} +\frac{V'-2V/\Psi}{24 \pi} \,
\end{split}
\ee
The right side of this equation can be non-negative if its determinant is less than zero. This translates to the condition
\be
\Theta^2 < \frac{4}{3} \; (2\frac{f(R)}{f'(R)} - R)
\ee
Thus, if the expansion scalar for a given geometry obeys the above inequality for a given $f(R)$ theory, then the gravity theory necessarily diverges away from GR. We can analyze the $V'-\frac{2V}{\Psi}$ combination further in the Jordan frame but to obtain a consistent criterion on the underlying f(R) theory (to distinguish whether it heats or cools gravity), we have to impose non-trivial constraints on the theory which are not very well physically motivated. So, for the case at hand, we instead switch over to the Einstein frame through a conformal transformation \cite{odinstovconformal} which makes us rely on only the constraints which ensure stability of the f(R) theory i.e . $f'(R)>0, f''(R)>0$ \cite{stabilityfaraoni,subhamssg}. But, then we shall focus on the chemical potential dynamics instead of temperature (which is trivially zero in this frame) \cite{pujolas2011imperfect,faraoni6},
\begin{equation}
    \frac{d\mu}{d\tau} = -\mu \Theta + \square\Phi
\end{equation}
where $\mu = \sqrt{-\nabla^a\Phi \nabla_a\Phi}$, $\square \Phi = V'(\Phi)$ and $\Phi$ is the Einstein frame (minimally coupled) scalar. $V'(\Phi)$ turns out to be $\frac{1}{\sqrt{3}}[Rf'(R) - 2f(R)]$. Thus, from the prior equation, we see that $Rf'(R) - 2f(R) < 0$ to help the theory tend towards GR, which essentially boils down to $\frac{d}{dR}\ln(\frac{f(R)}{R^2})<0$. In other words, $f(R)$ has to grow slower than $R^2$ to aid the attractor mechanism. For a Starobinsky theory $f(R) = R+\gamma R^2$, this attractor-to-GR condition reduces to having positive curvature theories ($R>0$). We should also notice that pure quadratic gravity $f(R) \propto R^2$ neither helps nor hinders the gravity theory to converge to GR.\\

\noindent \textit{Discussion\textemdash} We have provided a consistent thermal description of scalar-tensor effective theories in braneworld scenarios emphasizing the attractor mechanism towards GR. We explored the two-brane Randall-Sundrum model where the low-energy theory on either brane is of scalar-tensor nature with the extra-dimensional radion playing the role of the Brans-Dicke scalar. We further explored the detunings of the brane tensions in context of a non-vanishing gravitational contribution to radion potential, and found that in the \enquote{almost flat Planck brane} and the \enquote{almost flat TeV brane} regimes, non-dynamical detunings of the brane tensions exactly determine the signature for the radion potential contribution to the ${\cal KT}$ dynamics of the gravity theory (whether it \enquote{cools} or \enquote{heats} gravity) on either brane. We also identified a case where simultaneous detuning of the brane tensions could result in a \enquote{cooling}(\enquote{heating}) contribution without adhering to a regime if the Planck brane geometry is of the anti-de Sitter (de Sitter) type. We then discussed that the radion potential contribution by a bulk Goldberger-Wise scalar in fact \enquote{cools} down gravity for a positive mass squared scalar (which is known to stabilize the radion) and \enquote{heats} gravity for a negative mass squared scalar (which fails to stabilize the radion). Restricting to fluctuations about the stabilized radion (upon including a bulk stabilizing field), we showed that the critical curves get smoothened to straight lines and that the ${\cal KT}$ dynamics on either brane is equivalent upto a conformal factor.\\ 

\noindent One should note that our present analysis is by no means exhaustive, and
that there may be several interesting extensions - RS with a time-dependent stabilizing field \cite{timedepsumantassg}, the non-flat brane RS model \cite{SauryaDas_2008,nonflateffective}, $f(R)$ braneworlds \cite{f(R)braneworld,shafaq}, multi-brane extensions \cite{multibrane1,multibrane2} and other higher-dimensional modified gravity scenarios where the low-energy effective theory is either scalar-tensor or more generally, satisfies the Eckart-Fourier constitutive equation \cite{eckart}.\\

\noindent In this context, it would also be interesting to explore braneworld models like the Dvali-Gabadadze-Porrati (DGP) model \cite{dgp} whose effective theories are of the general Horndeski variety, which is known to have a first-order thermodynamic interpretation. Some of these will also be addressed in our future works.\\

\noindent \textit{Acknowledgement\textemdash} SB is supported through INSPIRE-SHE Scholarship by the Department of Science and Technology, Government of India. SB acknowledges helpful discussions with participants of the J.V. Narlikar Memorial Conference on Cosmology and Astrophysics, 2025 held at Visva-Bharati, Santiniketan.

\appendix

\nocite{*}

\bibliography{trueref}

\end{document}